\documentclass[11pt,a4paper]{article}
\usepackage[hyperref]{nlp4MusA}
\aclfinaltrue
\usepackage{times}
\usepackage{latexsym}

\usepackage{url}
\usepackage{graphicx}
\usepackage{amssymb}
\usepackage{booktabs}
\usepackage{subfig}
\usepackage[justification=justified]{caption}

\title{Optimal Embedding Calibration for Symbolic Music Similarity}

\author{
    Xinran Zhang\textsuperscript{1}, 
    Maosong Sun\textsuperscript{1234}\thanks{\,   Corresponding author.}, Jiafeng Liu\textsuperscript{1}, Xiaobing Li\textsuperscript{1}\\
  \textsuperscript{1}Department of Music Artificial Intelligence and Music Information Technology \\Central Conservatory of Music, Beijing, China \\
  \textsuperscript{2}Department of Computer Science and Technology, Tsinghua University, Beijing, China\\
  \textsuperscript{3}Institute for Artificial Intelligence, Tsinghua University, Beijing, China\\
  \textsuperscript{4}State Key Lab on Intelligent Technology and Systems, Tsinghua University, Beijing, China\\
  \texttt{zhangxr.wspn@gmail.com, sms@tsinghua.edu.cn}
  }

\date{}
\begin{document}
\maketitle
\begin{abstract}
In natural language processing (NLP), the semantic similarity task requires large-scale, high-quality human-annotated labels for fine-tuning or evaluation. By contrast, in cases of music similarity, such labels are expensive to collect and largely dependent on the annotator's artistic preferences. Recent research has demonstrated that embedding calibration technique can greatly increase semantic similarity performance of the pre-trained language model without fine-tuning. However, it is yet unknown which calibration method is the best and how much performance improvement can be achieved. To address these issues, we propose using composer information to construct labels for automatically evaluating music similarity. Under this paradigm, we discover the optimal combination of embedding calibration which achieves superior metrics than the baseline methods.
\end{abstract}
\section{Introduction}\label{sec:introduction}

Symbolic music research has benefited a lot from natural language processing (NLP) paradigm with Transformer architecture ~\cite{NIPS2017_7181} and powerful pre-training models such as BERT ~\cite{devlin-etal-2019-bert} and GPT-2 ~\cite{radford2019language}. Addressing symbolic music problems as language modeling can utilize classical NLP functionalities such as semantic similarity. However, traditional methods for semantic similarity \cite{reimers-gurevych-2019-sentence,zhang-etal-2020-unsupervised,li-etal-2020-sentence} all require datasets with large-scale, high-quality human-annotated labels. By contrast, in cases of music similarity, these labels can be expensive to collect. Annotator's artistic preferences will also severely bias these labels. Hence they are difficult to deploy for music similarity.

On the other hand, some baseline methods only rely on a pre-trained language model and do not require fine-tuning. The widely used baseline methods include using a single specific token embedding (e.g., [\texttt{CLS}] token in BERT) as the sentence embedding, or averaging the token embeddings from the last Transformer layer as the sentence embedding, and then calculating the cosine distance between sentences as a measure of their similarity. However, recent study has revealed that these baseline methods \emph{perform poorly}. Results by \citet{reimers-gurevych-2019-sentence} demonstrate that these baseline methods actually can not outperform the GloVe algorithm \cite{pennington-etal-2014-glove}.

We consider the embedding calibration technique \cite{li-etal-2020-sentence,mu2018allbutthetop}. Recent findings indicate that standard normalization (SN) and nulling away top-\emph{k} singular vectors (NATSV, \citealp{li-etal-2020-sentence,mu2018allbutthetop}) can significantly improve the performance of semantic similarity without requiring fine-tuning, and thus are frequently used as baseline methods. Regrettably, these studies do not include cases that demonstrate all feasible calibration combinations, for example, averaging more than two Transformer layers (\citealp{li-etal-2020-sentence}). As a result, the performance boundary for embedding calibration remains unknown.


\begin{figure*}[hbt!]
\centering
\includegraphics[width=6in]{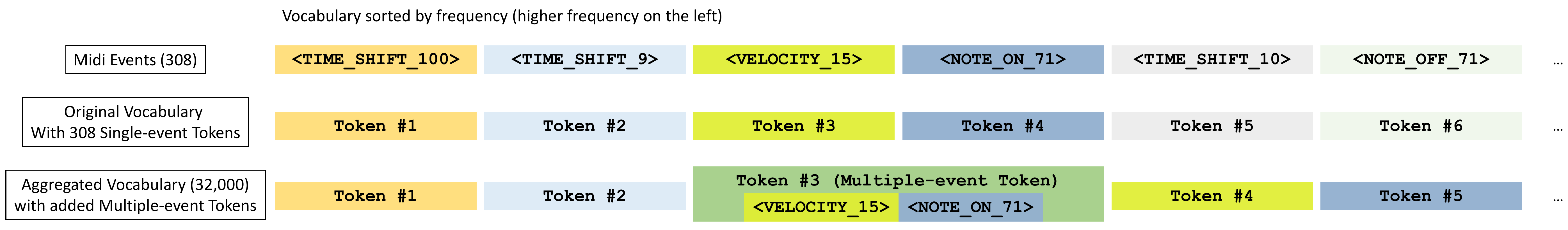}
\caption{Illustration of aggregated vocabulary with multiple-event tokens}\label{fig:vocab}
\end{figure*}



To address these issues, we investigate the optimal embedding calibration for music similarity. The following are our contributions.

\begin{itemize}
\item{We propose an automated method for evaluating music similarity that makes use of composer information. This method is capable of producing statistically significant results without requiring human annotations.}
\item{Under the proposed evaluation paradigm, we discover that the optimal performance is obtained by averaging the last 8 out of 12 Transformer layers, in combination with standard normalization calibration. The correlation metric increases to 0.223 when using the optimally calibrated embedding, compared to 0.154/0.028 when using baseline methods.}
\end{itemize}

\section{Embedding Calibration for Music Similarity}

\subsection{Pre-trained Language Model}

We release a pre-trained auto-regressive Transformer language model for symbolic music. To build the composer-centric labels, we choose the MAESTRO dataset \cite{hawthorne2018enabling} to train the model since the dataset contains large-scale, high-quality symbolic music data with accurate metadata. However, past works \cite{huang2018music,performance-rnn-2017} have used a vocabulary of only 308 words (midi events), which is far less than the vocabulary of benchmark NLP language models. Additionally, the music sequence is significantly longer than the typical language model's maximum context length. To remedy this, we employ a vocabulary aggregation technique, as illustrated in Figure \ref{fig:vocab}. Each midi-event in the vocabulary is treated as a single-length word. The Sentencepiece model \cite{Kudo_2018} is then used to aggregate and expand the vocabulary by integrating high-frequency single-event (single-length) words to multiple-event (multiple-length) words and adding them to the vocabulary. It will generate a vocabulary that includes both single-event and multiple-event words. The details of the pre-trained model are available in our publicly available code.

\subsection{Embedding Calibration}

Let $\mathbf{h}_{l,t}\in\mathbb{R}^{H}, l\in\{1,2,...,L\},t\in\{1,2,...,T\}$ denote the embedding vector of the token on layer $l$ and position $t$, where $L$ denotes the total number of Transformer layers, $T$ denotes the maximum context length and $H$ denotes the embedding size. We propose to investigate all possible combinations of the following three embedding calibration techniques.

\noindent \textbf{Last $\tilde{L}$ Layer Average Calibration:} averaging on the last $\tilde{L}$ Transformer layers, with $\tilde{L}\in\{1,...,L\}$.

\noindent \textbf{Standard Normalization (SN) Calibration:} $\mathbf{h}$ is calibrated by $(\mathbf{h}-\mathbf{\mu})/\mathbf{\sigma}$, where $\mathbf{\mu}$ and $\mathbf{\sigma}$ denote the mean and standard variance of all sentence embeddings.

\noindent \textbf{Nulling Away Top $\tilde{K}$ Singular Vector (NATSV) Calibration:} calculate and remove the first $\tilde{K}$ principal components in the sentence embedding space, with $\tilde{K}\in\{0,1,...\}$.

Note that the original NATSV algorithm \cite{mu2018allbutthetop} is calculated in the token embedding space. Conducting principal component analysis (PCA) on all tokens in the MAESTRO corpus will be challenging. We propose, heuristically, that PCA be performed on the sentence embedding space instead.
\begin{figure*}[hbt!]
\centering
\includegraphics[width=6in]{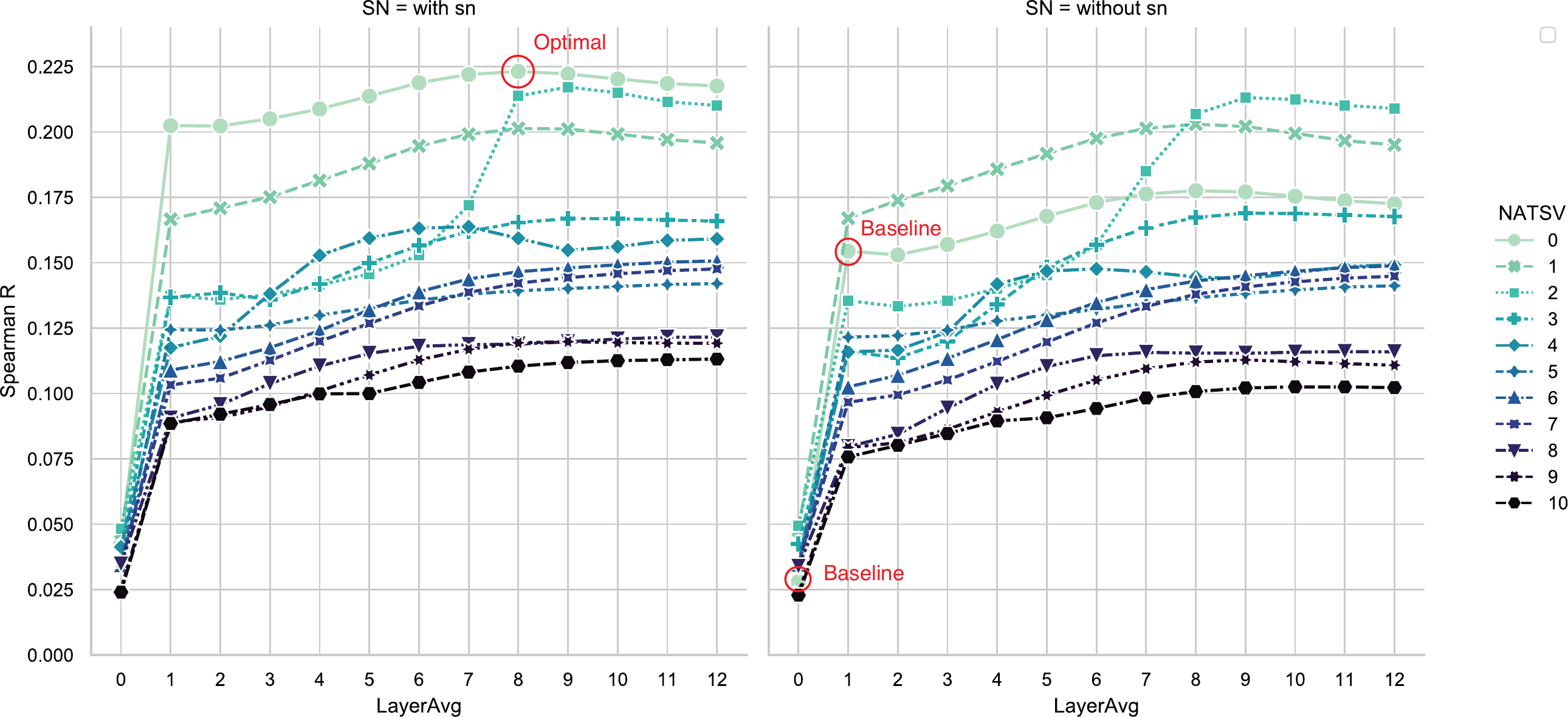}
\caption{Main results. The baseline methods (LayerAvg being 0 and 1) only achieve metrics of 0.028 and 0.154. The optimal metric (0.223) is achieved by averaging the last 8 layers with standard normalization calibration.}\label{fig:similarity}
\end{figure*}
\subsection{Automated Labels for Evaluation using Composer Information}

We propose an automated method for creating labels for the purpose of evaluating music similarity. It is worth noting that the MAESTRO dataset contains precise midi events that record performance data. By intuition, the internal consistency of music composed by the same composer is often more significant than the internal consistency of music composed by different composers. As a result, music sequences by the same composer will be more similar than sequences by different composers in general. Although this is not a perfect similarity annotation and hence cannot be used for fine-tuning, it may be used to evaluate all potential calibration methods fairly. To ensure statistical significance, the number of such labels can be much greater than the human-annotated NLP datasets (as in the cases in \citealp{reimers-gurevych-2019-sentence}).

We construct music sequences using a sliding window with a size equal to the maximum context length and a stride half the size of the window. Then we randomly select 100,000 pairs of sequences by the same composer and label them positively, and another 100,000 pairs of sequences by different composers and label them negatively. The similarity between two sequences can be measured using the cosine distance between their sentence embeddings. Then, as an evaluation metric for the calibration, we use the commonly used Spearman's correlation between the calculated similarity and the constructed labels. Without requiring a human-annotated dataset, we may now search for the optimal combination of embedding calibration.

\section{Experiments}

\subsection{Optimal Embedding Calibration}

The main results are shown in Figure \ref{fig:similarity}. ``LayerAvg'' on x-axis indicates $\tilde{L}$, and different markers indicates $\tilde{K}$. For simplicity of narration, we also refer to $\tilde{L}$ as ``LayerAvg value'', and refer to $\tilde{K}$ as ``NATSV value''. The left figure column indicates ``with standard normalization (SN)'', and the right figure column indicates otherwise. All experiments are statistically significant (with $p$-value being less than $10^{-4}$). Since our model is auto-regressive, the last token will have a full attention range that covers all tokens in the sequence. As a result, we choose the last token embedding as the baseline that corresponds to the [\texttt{CLS}] token embedding baseline, which is marked with LayerAvg being 0 on both figure columns. The other baseline is the points with LayerAvg being 1.
\begin{figure*}[hbt!]
\centering
\captionsetup[subfigure]{justification=centering}

\subfloat[Results of linear position weighting.]{\includegraphics[width=6in]{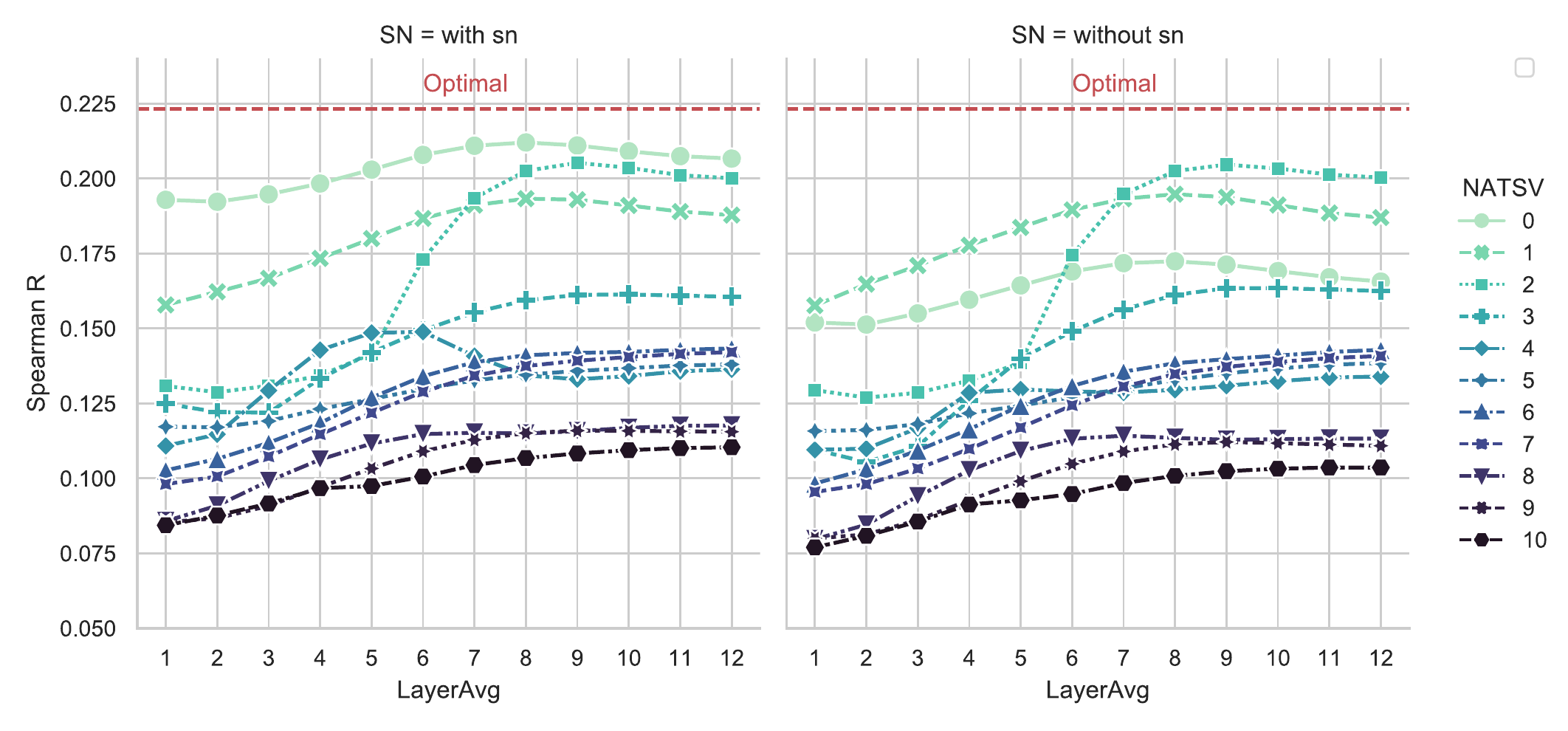}}\hfill
\subfloat[Results of inverse linear position weighting.]{\includegraphics[width=6in]{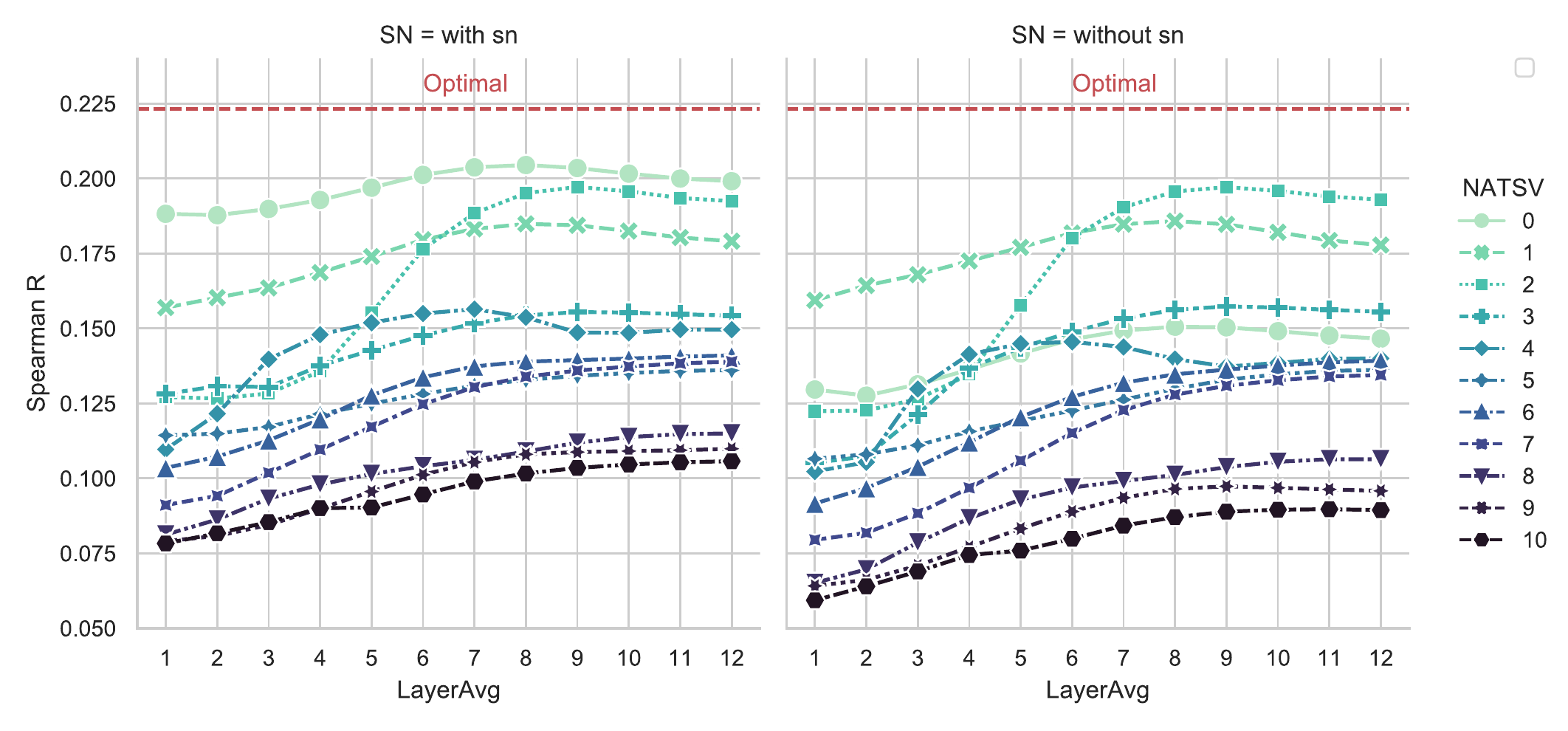}}\hfill

\caption{Impact of position weighting. It shows that linear position weighting outperforms inverse linear position weighting in terms of the correlation metrics.}\label{fig:similarity_pw}
\end{figure*}

As is shown in Figure \ref{fig:similarity}, the optimal metric is obtained on the left column by setting LayerAvg to 8, with SN, and setting NATSV to zero. Note that the correlation metric achieves its maximum value when LayerAvg equals 8. This demonstrates that while averaging more layers, as demonstrated by \citet{li-etal-2020-sentence}, is beneficial, there is an optimal point over which the correlation metric degrades.

%
%


The performance boost provided by SN is significant. The curve with SN and NATSV set to 0 in the left figure column is substantially better in Figure \ref{fig:similarity}. However, the gain with NASTV is only noticeable in a few spots and is not as large as the gain with SN. For instance, for the curve without SN in the right figure column, NASTV equals 1 is always preferable to 0, whereas NASTV equals 2 is only partially preferable to 1.


\subsection{Impact of Token Position}



%

Since our model is auto-regressive, which means that the attention span of each token is different than that of the auto-encoder Transformer model, a further question arises: How does the position of the token affect its importance for similarity? We undertake the following experiment to address this question. Given the fact that the current method of calculating sentence embedding takes the average of all positions, which implies that all positions are equally important, we examine two scenarios in which the token's importance is unequal and proportional to its position.

\noindent \textbf{Linear Position Weighting:}
\begin{equation}
\mathbf{h}_{l}=\bigg(\sum_{t=1}^{T}{t\times\mathbf{h}_{l,t}}\bigg)/\sum_{t=1}^{T}{t},
\end{equation}
%
%
\noindent \textbf{Inverse Linear Position Weighting:}
\begin{equation}
\mathbf{h}_{l}=\bigg(\sum_{t=1}^{T}{(T-t+1)\times\mathbf{h}_{l,t}}\bigg)/\sum_{t=1}^{T}{t},
\end{equation}
where $\mathbf{h}_{l}$ is the intermediate embedding computation for layer $l$, which is subsequently averaged across layers to obtain the sentence embedding $\mathbf{h}$. Results are shown in Figure \ref{fig:similarity_pw}. They demonstrate that linear position weighting outperforms inverse linear position weighting in terms of the correlation metrics. Additionally, the former can obtain an optimal metric that is fairly close to Figure \ref{fig:similarity}, whereas the latter drops more. These results indicate that in an auto-regressive Transformer model, tokens with a larger position index in the sequence are more important for music similarity.

\section{Conclusion}

We investigate music similarity using embedding calibration. We propose an automated method for constructing labels using composer information for evaluation. Our results show that the optimal embedding calibration is obtained by averaging the last 8 out of 12 layers of token embeddings using standard normalization calibration. The correlation metric significantly increases compared to the baseline methods. This method only requires the pre-trained language model and does not rely on additional datasets for fine-tuning.
\bibliographystyle{nlp4MusA_natbib}
\bibliography{poetry2020}

\end{document}